\documentclass[%
 reprint,
superscriptaddress,
 amsmath,amssymb,
 aps,
 prl,
]{revtex4-2}

\usepackage{soul}
\usepackage{graphicx} 
\usepackage{hyperref}
\usepackage{dcolumn}
\usepackage{bm}
\usepackage[mathlines]{lineno}

\usepackage{xcolor}
\usepackage[normalem]{ulem}

\begin{document}



\title{An estimate for thermal diffusivity in highly irradiated tungsten using Molecular Dynamics simulation}

\author{Daniel R. Mason}%
\affiliation{UK Atomic Energy Authority, Culham Science Centre, Oxfordshire OX14 3DB, UK}%
\email{Daniel.Mason@ukaea.uk}

\author{Abdallah Reza}
\affiliation{Department of Engineering Science, University of Oxford, Parks Road, OX1 3PJ, UK}
\email{mohamed.reza@eng.ox.ac.uk}

\author{Fredric Granberg}
\affiliation{Department of Physics, University of Helsinki, P.O. Box 43, FI-00014, Helsinki, Finland}%

\author{Felix Hofmann}
\affiliation{Department of Engineering Science, University of Oxford, Parks Road, OX1 3PJ, UK}
\email{felix.hofmann@eng.ox.ac.uk}

\date{June 2021}    
 
\begin{abstract}
The changing thermal conductivity of an irradiated material is among the principal design considerations for any nuclear reactor, but at present few models are capable of predicting these changes starting from an arbitrary atomistic model. Here we present a simple model for computing the thermal diffusivity of tungsten, based on the conductivity of the perfect crystal and resistivity per Frenkel pair, and dividing a simulation into perfect and athermal regions statistically. This is applied to highly irradiated microstructures simulated with Molecular Dynamics. A comparison to experiment shows that simulations closely track observed thermal diffusivity over a range of doses from the dilute limit of a few Frenkel pairs to the high dose saturation limit at 3 displacements per atom (dpa).
\end{abstract}

\maketitle

\section{Introduction}

Tungsten has been chosen as a plasma facing material designs for future tokamak fusion reactors~\cite{You2016,Donne2018,Hirai2016} due to its low sputtering yield, high melting point and high thermal conductivity~\cite{Rieth_JNM2011}. 
But under bombardment from 14.1 MeV fusion neutrons, displacement damage within the bulk material will generate lattice defects~\cite{Fukuda_JNM2014} which can adversely affect thermal conductivity among other properties~\cite{Hasegawa_JNM2011}.

Unfortunately, predicting thermal conductivity based on the damage microstructure is extremely difficult, as metal conductivity is dominated by electrons, and so requires a quantum mechanical treatment.
The electron scattering rate can be written down from Fermi's golden rule as proportional to the square of a perturbing matrix element coupling two electron states.
For the electron-phonon coupling this can be computed from the elastic deformation due to the phonon~\cite{Khan_PRB1984}.
In semiconductors at least sufficient electron localisation is present to permit fast scaling methods using Density Functional Perturbation Theory~\cite{Ganose_Nature2021}.
Time-dependent tight binding has also been used to find electron conductivity across molecules and nanowires with open boundaries~\cite{Horsfield_PRB2016}.
These calculations are generally expensive and while transport calculations can be performed in the Boltzmann theory approximation~\cite{Madsen_CPC2006}, and scattering rates can be found~\cite{Alfred_PhysRev1966,Gupta_PRB1979,Gupta_PRB1987,Brandbyge_PRB2002}, current state-of-the-art ground-state density functional calculations of dislocation loops are limited to order one thousand atoms~\cite{Domain_JNM2018}. 
When this scale is compared to the minimum size for generating high dose microstructures, order one million atoms~\cite{Derlet_PRM2020}, we must concede that electronic structure calculations must be supplemented by more approximate methods if a fully multiscale picture of a material's response to stress, temperature and irradiation is to be developed. 

This simplifying approach was followed by Zinkle (ref~\cite{Zinkle_JPhysF1988}), who suggested a model for the resistivity of circular dislocation loops in copper based on counting defected atoms observed in TEM images and dividing these into dislocation core sites and atoms in stacking fault sites. 
Reza et al.~\cite{Reza_Acta2020} considered similar models, again based on TEM observations of atoms. 
It is noteworthy that both these papers required an extrapolation of the distribution of observed loops to sizes too small to observe~\cite{Liu_Acta2017,Mason_EPL2018}.
Caturla et al.~\cite{Caturla_JNM2001} modelled resistivity changes during post irradiation annealing using the resistivity per Frenkel pair, following the count of pairs using kinetic Monte Carlo.

We argue that to predict a thermal conductivity for engineering purposes it is sufficient to be able to divide an arbitrarily complex, atomically-detailed simulated microstructure into regions which are essentially perfect crystal, regions which are elastically distorted and so are somewhat scattering, and regions which are highly distorted and have substantially greater scattering. 
If we can robustly predict and characterize an irradiated material along these lines, and reproduce the scattering rates of simple defect types, we should be able to reproduce the trends in conductivity change due to irradiation dose, temperature, stress and other external drivers through their effect on the microstructure, even if the scattering rate for an individual complex defect type is not exactly reproduced.

Existing methods for distinguishing athermal atoms from bulk crystal atoms include analysing bond angle distributions, common neighbour analysis and graphs of connected bonds~\cite{Ackland_PRB2006,Bhardwaj_CompMatSci2020}.
Progress has also been made recently to detect athermal atoms based on Machine Learning~\cite{Goryaeva_NatComm_2020}.
We distinguish perfect lattice from distorted using local potential energy- a property generally available using empirical potentials even if not well-defined in an ab-initio calculation. 
This choice is made because we can derive an expression for the expected \emph{distribution} of atomic potential energy for a system in thermal equilibrium, combining the Maxwell-Boltzmann distribution with the Debye-Waller factors for thermal vibrations. We demonstrate that this distribution is a very good fit to MD simulations.

We then use a simple model for the electron scattering rate based on Mattheisen's rule~\cite{Mott_1936} for summing rate contributions on an atom-by-atom basis.
We use an empirical model for the scattering rate due to an atom in a defected configuration~\cite{Mason_JPCM2015}, and describe how to parameterize an empirical potential to fit thermal conductivity quantities using the scattering rate per Frenkel pair- a number which has been experimentally determined for many metallic elements.
With this model, we can uniquely define the thermal conductivity of arbitrarily complex atomic configurations.

The total thermal conductivity also has a component due to the phonons. Typically the phonon conductivity of metals is order 2-18 W/mK \cite{Tong_PRB2019}, significantly smaller than the electronic contribution. Tungsten's phonon contribution at room temperature has been estimated from molecular dynamics at 15-16 W/mK \cite{Hu_APL2017}, an order of magnitude smaller than the total thermal conductivity (174 W/mK) \cite{Ho_JPhysChemRefDat1972}. The phonon contribution decreases with both temperature and the number of irradiation-induced defects- mirroring the electronic contribution. For this work it is therefore possible to find the thermal conductivity assuming it is electronic only in origin, and ignore the small correction due to phonons. For other metals the validity of this assumption should be tested, and we discuss how to add the phonon contribution below.

Finally we compare the computed thermal diffusivity of simulated high dose tungsten microstructures, and compare to experimental measurements of high-dose self-ion irradiated tungsten with matching elastic boundary conditions. We show a very high quality match between the two. Importantly our simulated results are a much higher fidelity match than an estimate based unrelaxed high dose microstructures. This gives us confidence that our model is not just finding an order-of-magnitude estimate, but is tracking the variation of thermal conductivity as microstructure evolves.

\section{Theory}
    \label{sec:theory}
We can write a simple kinetic theory expression for the electronic thermal conductivity,
    \begin{equation}
        \label{eqn:kinetic_theory}
        \kappa_{el} = \frac{1}{3 \Omega_0} c_e v_F^2 \langle r_e \rangle^{-1},
    \end{equation}
where $c_e$ is the electronic heat capacity per atom,$\Omega_0$ is the atomic volume, $v_F$ is the Fermi velocity, and $r_e$ is the electron scattering rate.
The heat capacity is given in terms of the temperature $T$ and density of states at the Fermi level $D_F$, $c_e = (\pi^2 k_B^2 D_F/3) T$.

Electron scattering comprises contributions from impurity scattering, electron-phonon scattering and electron-electron scattering, with the condition that the electron mean free path cannot drop below the nearest neighbour separation $b_0$~\cite{Mason_JPCM2015}. 
    \begin{equation}
        \label{eqn:scattering_rate}
        \frac{1}{r_e} = \frac{b_0}{v_F} + \frac{1}{r_{imp} + r_{e-ph} + r_{e-e}}.
    \end{equation}
We expect impurity scattering to arise from electrons scattering from the anomalous electrostatic potential at defected sites, impurity atoms and the like, and so be temperature independent.
Electron-phonon scattering should be proportional to the number of phonons, and so scale linearly with $T$.
Finally electron-electron scattering should scale with $T^2$.
It is beyond the scope of this work to seek analytic expressions for the latter two terms, so instead we fit to the known variation of thermal conductivity with temperature, and write $r_{e-ph} = \sigma_1 T$, and $r_{e-e} = \sigma_2 T^2$~\cite{Baber_PRSA1936}. 
We note that this implies our model has an unphysical infinite conductivity for the perfect lattice at zero temperature; in reality there will always be some residual defects and scattering between s- and d- bands in transition metals~\cite{Tsiovkin_PRB2005}, but resistivity ratios $\rho(273$K$)/\rho(4.2$K$)$ of order 10$^5$ can be measured for very pure single crystal tungsten samples~\cite{Rasch1980}.

In this work we focus on the impurity scattering. The experimental literature for scattering rates for specific defects is sparse, owing to the difficulty of knowing exactly which defects are present, but we summarise three important results. 
In ref~\cite{Shukovsky_Acta1966}, the electrical resistivity per vacancy in tungsten was observed to be proportional to linear strain. 
    Secondly, if the resistivity per Frenkel pair~\cite{Broeders_JNM2004} is compared to the resistivity per vacancy~\cite{Ullmaier_1991} for molybdenum and tungsten, we find similar ratios of 3.1 and 3.9 respectively. 
    Thirdly, in ref~\cite{Alfred_PhysRev1966}, the resistivity for point defect pairs in copper ( divacancy and di-interstitial ) is calculated to be slightly under double the single point defect value, consistent with best estimates from experiment. 
    These three results suggest that the defect scattering rate correlates with excess energy: 
    the formation energy per vacancy is expected to vary linearly with strain, with the (tensorial) coefficient being the dipole tensor~\cite{Dudarev_Acta2017}.
    The formation energy ratios of Frenkel pair to vacancy computed by DFT ( using AM05 potential ) for Mo and W are 3.5 and 4.0 respectively~\cite{Ma_PRM2019}, which is a reasonable fit to the second observation.
    The third observation would be consistent with a small binding energy for point defects.
We therefore suggest an empirical model, $r_{imp} = \sigma_0 |E|$, where $E$ is the excess potential energy of a defected atom~\cite{Mason_JPCM2015,Hofmann_SciRep2015}. Note that we use the modulus to prevent unphysical negative rates; in practice few defected atoms have negative excess energies, so for the purposes of exposition it is convenient to assume scattering rate from a defect at low temperature is proportional to its formation energy.
How we define excess energy, and whether an atom is defected or not is given below.

Consider a system of atoms thermalized using classical molecular dynamics at temperature $T$ with an empirical many body potential.
The energy $E$ in a particular phonon mode with frequency $\omega$ is given by the Boltzmann distribution, $p_B(E;T) dE = \beta \exp[ - \beta E] dE$, where $\beta = 1/k_B T$ is the inverse temperature.
From this, it is straightforward to show that the kinetic energy of each atom follows the Maxwell-Boltzmann distribution, $p_{M-B}(E;T) dE =  \beta (2 \beta E)^2 \exp[ - 2 \beta E] dE$.
The potential energy of each atom does not quite follow this distribution, as the atoms are not Einstein oscillators but rather have local energies determined by the distances to their neighbours.
But if we assume that for thermally equilibrated atoms, they nevertheless appear to be close to Einstein oscillators, it follows that the probability distribution of the position of each atom is close to a spherically symmetric Gaussian. This approximation is often used in constructing Debye-Waller factors for dynamical electron diffraction calculations: the Debye-Waller factor, $B$, is related to the thermally averaged atom displacement in the $x-$ direction, $B=8\pi^2 \langle u_x^2 \rangle$, where in the harmonic approximation~\cite{Lovesey_1984},
    \begin{equation}
        \langle u_x^2 \rangle = \left(\frac{\hbar}{2m}\right) 
        \int \mathrm{coth} \left(\frac{\hbar \omega}{2 k_B T}\right) \, \frac{g(\omega)}{\omega} \mathrm{d}\omega,
    \end{equation}
with $g(\omega)$ being the normalised phonon density of states.
We can find the temperature scaling of this displacement scale by using the Debye formula in place of the density of states, to give\cite{Peng_ActaCryst1996}
    \begin{equation}
        \langle u_x^2 \rangle = \left( \frac{11492}{8 \pi^2 M} \right) \left(\frac{T}{\Theta_D^2} \right) \left( \Phi \left(\frac{\Theta_D}{T}\right) + \frac{1}{4} \left(\frac{\Theta_D}{T}\right) \right),
    \end{equation}
where $\Theta_D$ is the Debye temperature and $\Phi(\Theta_D/T)$ is the Debye integral. 
If $M$ is the atomic mass in Daltons, then $\langle u_x^2 \rangle$ is returned in units of $\AA^{2}$.
Above the Debye temperature (or in classical molecular dynamics where quantum mechanical phonons are not represented), $\langle u_x^2 \rangle$ scales linearly with $T$, and so
    \begin{equation}
        \langle u_x^2 \rangle \sim \frac{145.55}{M \Theta_D^2}  T.
    \end{equation}
For tungsten, $\Theta_D= 312$ K \cite{Ho_JPhysChemRefDat1972}.

With this Gaussian approximation for atom positions, the probability distribution for the distance between the atoms must \emph{also} be Gaussian, albeit with a slightly larger half-width of the distribution, $w$. If the perfect lattice distance between atoms is $R^{(0)} \gg w$, then the probability distribution at finite temperature is
    \begin{equation}
        p(R) \approx \frac{1}{\sqrt{2 \pi w^2}} \exp \left( - \frac{(R-R^{(0)})^2}{2 w^2} \right),
    \end{equation}
with $w^2 = 16 \langle u_x^2 \rangle/\pi^2$. 
Hence we can say that the standard deviation of the bond-length fluctuations scales as $w \sim \sqrt{T}$.

As we are assuming the thermal vibrations are small, we can linearise the energy dependence in terms of atomic separations, and so find the probability distribution for potential energies will be approximately given by the convolution of the Maxwell-Boltzmann distribution and a broadening function, $g(E;\sigma) = \exp[ -E^2/(2 \sigma^2) ]/\sqrt{ 2 \pi \sigma^2 }$. The preceding arguments suggest that $\sigma^2\sim \Delta k_B T$, with $\Delta$ a potential dependent constant with energy units. 
We shall see below this energy parameter is easily found from simulation.
With the convolution applied, we find our form for the distribution of potential energies in a thermalised MD simulation:
    \begin{eqnarray}
        \label{eqn:potential_energy_distribution}
        p_{\rm{MD}}(E;T) &=& p_{M-B}(E;T) \otimes g(E;\sigma)        \nonumber   \\
        &=&
        2 \beta^3 \left\{ \exp\left[ -\frac{E^2}{2\sigma^2} \right] 
        \sqrt{ \frac{2 \sigma^2}{\pi} } (E - 2\beta \sigma^2 )  \right.    \nonumber\\
        && \quad  + \exp\left[ 2 \beta^2 \sigma^2 - 2\beta E\right] (\sigma^2 + (E-2 \beta \sigma^2)^2 ) \nonumber\\
        &&
        \quad\quad\quad \left. \times( 1 + \mathrm{erf}\left( \frac{ E - 2 \beta \sigma^2 }{\sqrt{ 2 \sigma^2 }} \right) \right\} \nonumber \\
    \end{eqnarray}
The zero of energy is taken here to be the energy per atom at zero temperature with appropriate supercell strains applied, and so $E$ is the excess potential energy.

\begin{figure}
    \centering
    \includegraphics[width=0.7 \linewidth,angle=-90] {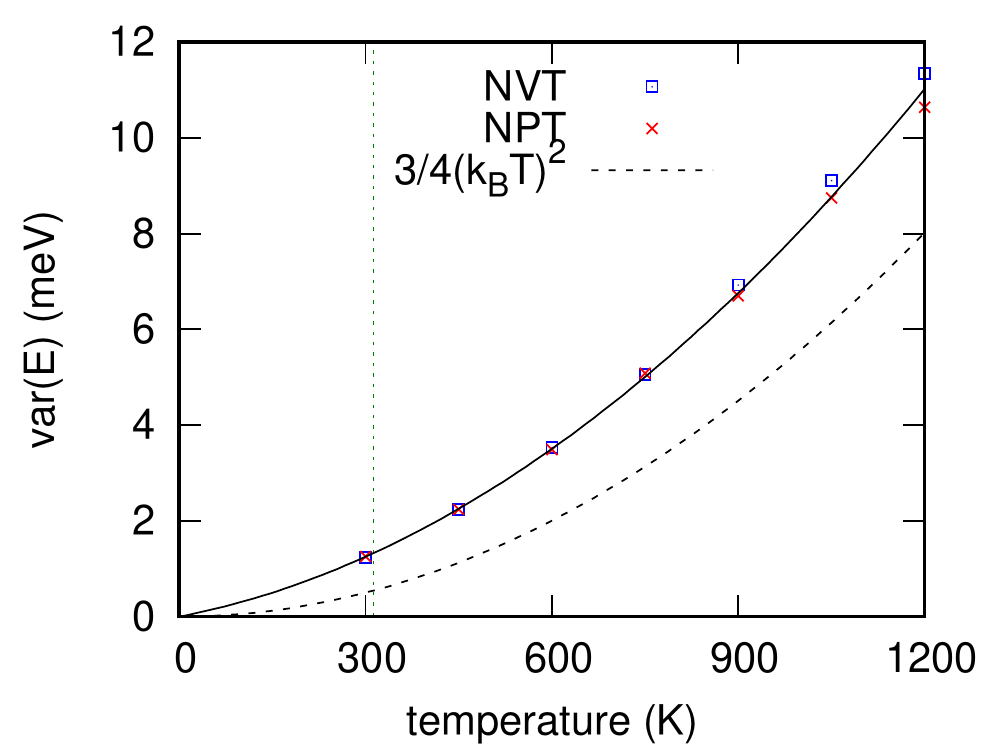}
    \caption{The variance of the potential energy of 65k atoms  thermalized in the NVT and NPT ensembles. The dashed line shows the variance in the Maxwell-Boltzmann distribution, and the solid line is the model including broadening (equation \ref{eqn:potential_energy_distribution}) with $\sigma = \sqrt{ \Delta (k_B T) }$, with $\Delta = 0.029$ eV.
    The vertical line shows the position of the Debye temperature in tungsten, $\Theta_D = 312$ K.
    }
    \label{fig:variance}
\end{figure}

\begin{figure}
    \centering
    \includegraphics[width=0.7 \linewidth,angle=-90] {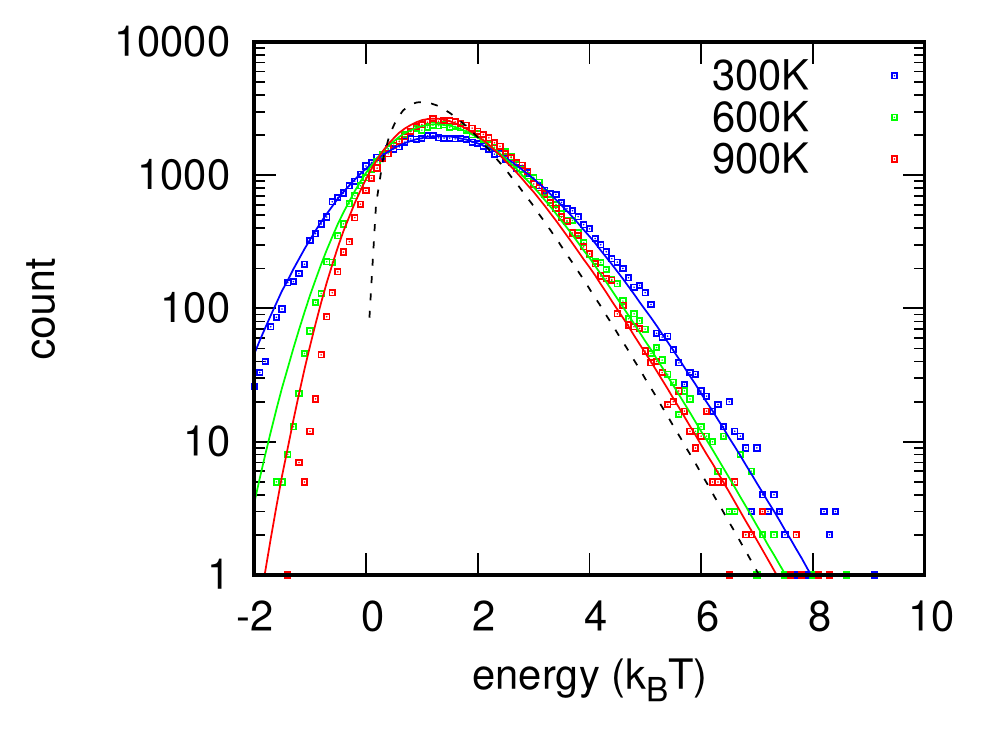}
    \caption{A histogram of potential energies of 65k atoms thermalized in the NPT ensemble using LAMMPS.
The dashed line is the M-B distribution, and the solid lines are a convolution with a Gaussian width $\sigma = \sqrt{ \Delta (k_B T) }$ (equation \ref{eqn:potential_energy_distribution}). 
    }
    \label{fig:broadenedMB}
\end{figure}

The first few moments of $p_{MD}(E;T)$ are:
$\int p_{\rm{MD}}(E;T) dE = 1$, $\int E p_{\rm{MD}}(E;T) dE = 3/2 k_BT$, and $\int E^2 p_{\rm{MD}}(E;T) dE = 3 (k_BT)^2 + \sigma^2$.
The simple form for the second moment means we can parameterize for $\sigma$ by plotting the variance of the potential energy as a function of temperature.
We thermalize a simulation box of $65336$ tungsten atoms using LAMMPS~\cite{LAMMPS} and an empirical potential~\cite{Mason_JPCM2017} known to give reasonable point defect and thermal expansion properties. 
In figure~\ref{fig:variance} we show that the variance is a good fit to the form $\mathrm{var}(E) = 3/4 (k_B T)^2 + \Delta k_B T$ in both NPT and NVT ( Number, Pressure or Volume and Temperature ) ensembles, and that in both ensembles $\Delta = 0.029 \pm 0.001$ eV.
The high quality of a broadened Maxwell-Boltzmann distribution is further shown in figure~\ref{fig:broadenedMB}.
Here we have generated a histogram of the potential energy per atom for the $65336$ atom box in the NPT ensemble.
Note that the fit is good even in the tails of the distribution.

If we generate a histogram of potential energies similar to figure~\ref{fig:broadenedMB} but in a defected system of atoms, and compare to the expected thermal distribution (equation~\ref{eqn:potential_energy_distribution}), we can estimate how many atoms are thermal, and how many are athermal.
Note that we can not say for certain whether an individual atom is defected, only find the fraction of athermal atoms in each energy bin.

\begin{figure}
    \centering
    \includegraphics[width=0.7 \linewidth,angle=-90] {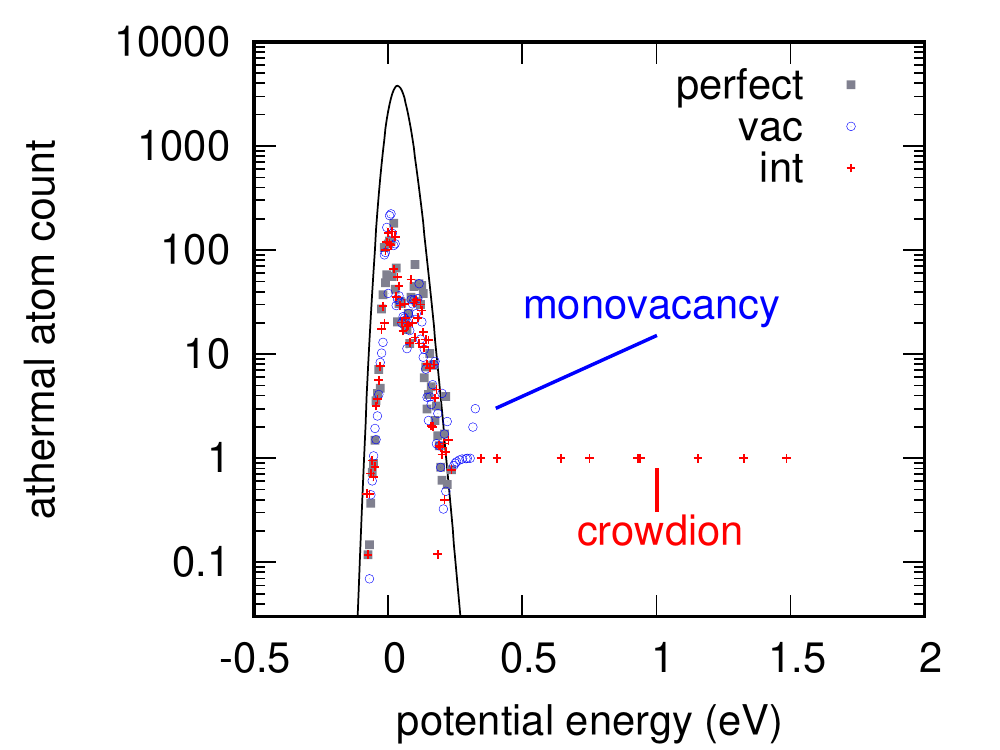}
    \caption{A histogram of athermal atoms in a system of 65k atoms  thermalized in the NPT ensemble at 300K.
    The solid line shows the expected count of thermal atoms in each bin, two orders of magnitude higher than the athermal count.
    The symbols show the predicted number of athermal atoms (equation \ref{eqn:defect_count}, for a defect-free lattice, and for monovacancy crowdion configurations.
    }
    \label{fig:pointDefects}
\end{figure}

If there are $N$ atoms total in the system, then we expect to find a number $\bar{N}$ in the energy range $E:E+\mathrm{d}E$ given by $\bar{N}(E;T) =  N\, p_{\rm{MD}}(E;T) \mathrm{d}E$.
The actual number of thermal atoms we record should follow a Poisson distribution with this average, ie the distribution $\Pi(n;\bar{N}) = \bar{N}^n \exp[ -\bar{N} ] / n!$
~\footnote{Strictly speaking, the number of atoms in each energy window must be correlated if the total count is fixed. But in the limit of a large number of atoms and many bins, this correction becomes negligible.}.
If we actually record $n$ atoms in the energy interval, then the probability that $k$ of these are non-thermal atoms must be given by the Poisson probability that $n-k$ are thermal
    \begin{equation}
        p(k ; n,\bar{N} ) = \frac{ \Pi(n-k;\bar{N}) } { \sum_{k=0}^n \Pi(n-k;\bar{N}) }.
    \end{equation}
The expected number of non-thermal atoms in this energy window is therefore
    \begin{equation}
        \label{eqn:defect_count}
        \langle k \rangle = \sum_{k=0}^n k \, p(k ; n,\bar{N} ).
    \end{equation}
Histograms of athermal atom count using equation~\ref{eqn:defect_count} for systems containing a single point defect are shown in figure~\ref{fig:pointDefects}.
Note that the expected number of non-thermal atoms defined in this way tracks the thermal count, simply because this is a stochastic property of the system. ( The athermal proportion is order 2\% for this potential and system size, a value largely independent of temperature ). The true signal of the point defects appears where we expect to see very few thermal atoms. For the monovacancy at 300K, we see a signal at 0.3eV. This is generated by the cage of high energy atoms surrounding the vacancy itself. For the crowdion we see the individual atoms making up this extended defect with very high energy ($>0.5$ eV). 

We can compute expected scattering rates for thermal atoms using equation \ref{eqn:scattering_rate}:
    \begin{equation}
        r_{\theta}(T) = \frac{v_F (\sigma_1 T + \sigma_2 T^2 )  }{b_0 (\sigma_1 T + \sigma_2 T^2 ) + v_F },
    \end{equation}    
and for athermal atoms with 
    \begin{equation}
        r_{i}(E;T) = \frac{v_F (\sigma_0 |E| + \sigma_1 T +\sigma_2 T^2 )  }{b_0 (\sigma_0 |E| + \sigma_1 T + \sigma_2 T^2 ) + v_F }.
    \end{equation}

We can therefore find the expected scattering rate due to electron-phonon and impurity scattering from atoms in the energy window $E:E+dE$ is
    \begin{equation}
        r(E;T) = \sum_{k=0}^n p(k ; n,\bar{N}(E;T) ) \, \left( (n-k)  r_{\theta}(T) + k r_{i}(E;T) \right),
    \end{equation}
and the total scattering rate is
    \begin{equation}
        \label{eqn:total_scattering_rate}
        r_e = \int r(E;T) dE.
    \end{equation}
In practice we need to generate a histogram, so this integral is computed numerically. 
The scattering rate is not biased by bin width provided the width is small compared with the temperature scale. We use bin widths $\mathrm{d}E \sim k_B T/20$.

\subsection{Fitting the model to experiment}
\label{sec:fitting}

In the limit $T\rightarrow 0$, all atoms in a perfect crystal have $E=0$. 
For a crystal containing a point defect relaxed using conjugate gradients no atoms will have exactly $E=0$, although most will be in a narrow bin $-\mathrm{d}E/2:+\mathrm{d}E/2$. Atoms outside this bin can be assumed `athermal' in the low temperature limit.
    
\begin{figure}
    \centering
    \includegraphics[width=0.7 \linewidth,angle=-90] {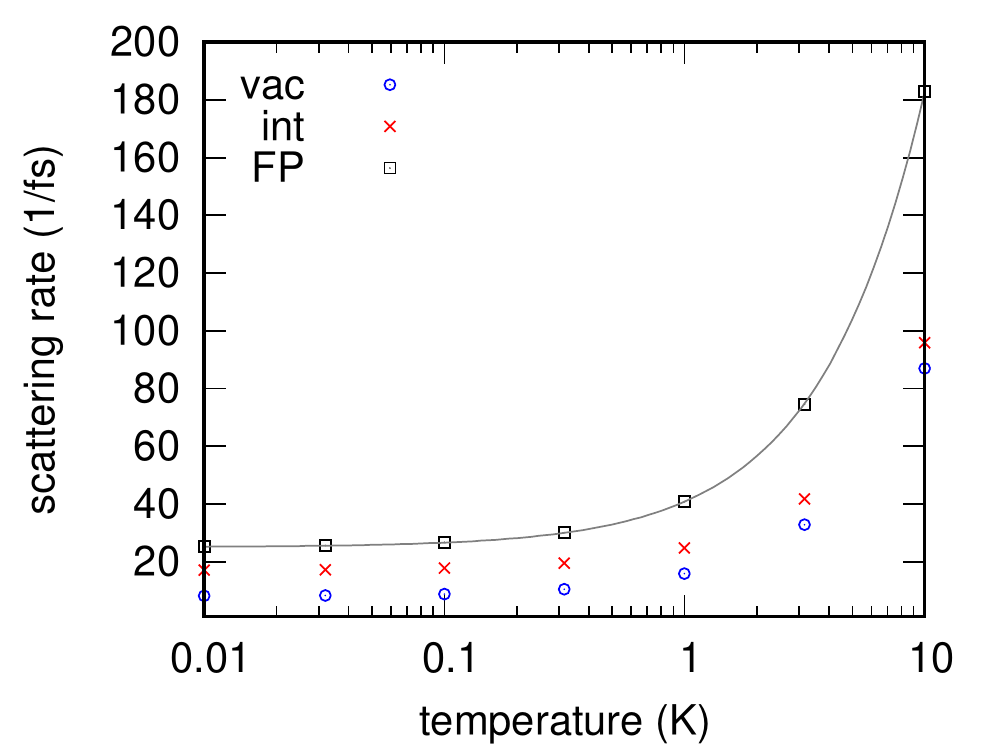}
    \caption{
    Scattering rate computed for monovacancy and crowdion point defects in perfect lattice, with assumed temperatures applied. The solid line is an affine fit, so the y-axis intercept gives the scattering rate for the Frenkel pair at $T=0$.
    }
    \label{fig:fitSigma0}
\end{figure}

We can compute scattering rate for a defect relaxed using conjugate gradients, assuming a small temperature $T$ were applied to avoid the singularity in the rate at $T=0$, provided we make some choice for the triplet $\{\sigma_0,\sigma_1,\sigma_2\}$. 
The scattering rate for a Frenkel pair, $r_{\rm{FP}}(T)$, is just the sum of the rates for monovacancy and crowdion.
We can then use the Wiedemann-Franz law relating electrical resistivity to thermal conductivity, $\rho = L T/\kappa$, where $L=2.44\times 10^{-8}$ W$\Omega$K$^{-2}$ is the Lorentz number.
At low temperature, the phonon heat capacity, and hence the phonon thermal conductivity scales as $T^3$ according to the Debye Law, and so we can neglect phonon contributions in this limit.
We can therefore match the defect scattering constant, $\sigma_0$ to the measured resistivity per Frenkel pair, $\rho_{\rm{FP}}$, by substituting equation \ref{eqn:kinetic_theory}:
    \begin{equation}
        \label{eqn:Wiedemann-Franz}
        \rho_{\rm{FP}} = \lim_{T\rightarrow 0} \frac{ 3 L \Omega_0 }{v_F^2 \left( c_e/T \right) } r_{\rm{FP}}(T).
    \end{equation}
As $\lim_{T\rightarrow 0} r_{\rm{FP}}(T)$ is linear in $\sigma_0$, we can use this to fit $\sigma_0$. 
Using $v_F = 9.5 \AA$/fs and $c_e/T/\Omega_0 = 5.46\times 10^{-10}$ eV/K$^2$/\AA$^3$, computed using Density Functional Theory~\cite{Mason_JPCM2015}, and the experimental value $\rho_{\rm{FP}} = 27$ $\mu\Omega$ m/at.fr.~\cite{Broeders_JNM2004}, we find a target value $r_{\rm{FP}}(T=0)=29.1$ fs.
Figure~\ref{fig:fitSigma0} shows the fitting of our model to this computed scattering rate per Frenkel pair, achieved by setting $\sigma_0=2.32$ fs$^{-1}$/eV.
The error in this value due to the non-linearity of the computed rates $r_{\rm{FP}}(T)$ is very much smaller than the uncertainty in $\rho_{\rm{FP}}$.

With $\sigma_0$ fixed by the Frenkel pair calculation, we can fit $\sigma_1$ and $\sigma_2$ to reproduce the experimental thermal conductivity~\cite{Ho_JPhysChemRefDat1972}.
\begin{figure}
    \centering
    \includegraphics[width=0.7 \linewidth,angle=-90] {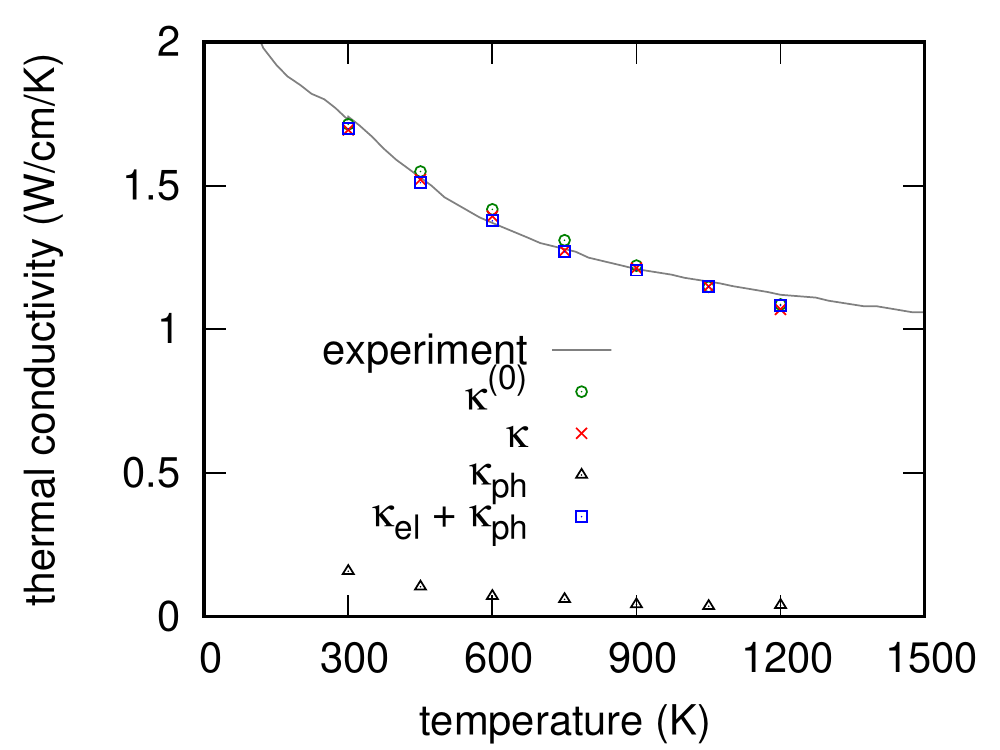}
    \caption{Thermal conductivity of atoms in a defect-free condition with three fitted sets of parameters for scattering: the perfect crystal lattice with an assumed temperature using the scattering coefficients $\sigma^{CG}_i$ ($\kappa^{(0)}$); thermalized in the NPT ensemble using the scattering coefficients $\sigma^{MD}_i$ ($\kappa$); and incorporating the phonon correction separately using the scattering coefficients $\sigma^{exc}_i$ and a Green-Kubo calculation of the phonon conductivity ($\kappa_{el}+\kappa_{ph}$).
    The phonon conductivity alone is also plotted separately.
    Solid line experimental data from ref~\cite{Ho_JPhysChemRefDat1972}.
    }
    \label{fig:diffusivity}
\end{figure}
Many methods to fit the curves would be appropriate here. 
We performed the fit efficiently by observing (empirically) that the fraction of athermal atoms is very weakly dependent on temperature, in these simulations $f(T) \approx 0.022 + 0.156 k_B T$, and their average energy is linear in temperature, $\langle E \rangle \approx 1.64 k_B T$. With these approximations we write down the expected scattering rate at temperature $T$ as
\begin{equation}
    \langle r (T) \rangle \approx f(T) r_i(\langle E \rangle;T) + (1-f(T)) r_{\theta}(T), \nonumber
\end{equation}
and hence the expected electronic thermal conductivity is 
    \begin{equation}
        \label{eqn:thermal_conductivity}
        \langle \kappa_{el} (T) \rangle  \approx  \frac{c_e v_F^2}{ 3  \langle r (T) \rangle }.
    \end{equation}
This is then a simple analytic form to fit for $\{\sigma_1,\sigma_2\}$.
We consider incorporating the correction due to phonon conductivity below.
The thermal conductivity predicted for a defect-free, but MD thermalized lattice at finite temperature is shown in figure~\ref{fig:diffusivity}.
We find a fit $\sigma^{MD}_1 = 1.154\times 10^{-4}$ fs$^{-1}$/K and $\sigma^{MD}_2 = 1.209\times 10^{-7}$ fs$^{-1}$/K$^2$.
The points in figure \ref{fig:diffusivity} for thermalized systems in the NVT and NPT ensembles use this set of parameters.
Note that the small change in homogenenous strain between the two ensembles does not significantly change the distribution of atomic energies ( except for their offsets ) at low temperatures, and so there is little difference in the calculated thermal conductivity in these ensembles in this temperature range.

In atomistic simulations we also often work with lattice statics, with relaxed atoms in their quasi-harmonic minima.
Just as we must take care not to treat thermal noise in atomic positions as genuine atomic disorder, so we must not treat the lack of noise in relaxed atomic configurations as an absence of disorder.
We can fit equation~\ref{eqn:thermal_conductivity} to the experimental data if the atoms are in ideal lattice positions. In that case we would expect no athermal atoms, ie a fraction $f=0$. 
This gives a fit which is suited to an atomic system which has been relaxed using conjugate gradients and has no thermal noise.
We find $\sigma_1^{CG} = 1.194\times 10^{-4}$ fs$^{-1}$/K and $\sigma_2^{CG} = 1.108\times 10^{-7}$ fs$^{-1}$/K$^2$. 
The points in figure \ref{fig:diffusivity} labelled as perfect crystal use this second set of parameters.
Note that $\sigma_1^{MD}$ is slightly smaller than $\sigma_1^{CG}$ as our statistical model always estimates a few percent of atoms in MD are `athermal' and so are given a higher scattering rate.
The closeness of the absolute values of $\sigma_i^{CG}$ and $\sigma_i^{MD}$ is an indication that harmonic vibrations are being correctly accounted for.

Note that in our model we ignore the contribution to thermal conductivity from phonons, which is computable using MD if needed, but here is small compared to electron conductivity.
Thermal diffusivity, $\alpha$, is defined from thermal conductivity as $\alpha = \kappa/c$,
where $c$ is the volumetric heat capacity, here dominated by phonons, so $c=3k_B/\Omega_0$.
A summary of the values used to parameterize and resultant conductivity is given for reference in table~\ref{tab:summary}.

\begin{table}[htb!]
    \centering
    \begin{tabular}{c|ccc}
        \multicolumn{4}{c}{Fitted parameters}\\
        \hline
                &   impurity            &   el-ph               &   el-el                   \\
                &   $\sigma_0$          &   $\sigma_1$          &   $\sigma_2$              \\
                &   fs$^{-1}$eV$^{-1}$  &   fs$^{-1}$K$^{-1}$   &   fs$^{-1}$K$^{-2}$       \\
CG-relaxed      &   2.32                &   $1.194\times10^{-4}$&   $1.108\times10^{-7}$    \\
MD              &   2.32                &   $1.154\times10^{-4}$&   $1.209\times10^{-7}$    \\
MD (exc $\kappa_{ph}$)     &   2.32                &   $1.344\times10^{-4}$&   $1.010\times10^{-7}$    \\ 
         \\
         \multicolumn{4}{c}{Derived properties} \\
         \hline
         broadening    &   $\Delta$      &   0.029 & eV    \\
         atom vol (T=0K)     &   $\Omega_0$  & 15.86 (15.86)$^{(a)}$  & \AA$^3$ \\
         conductivity & $\kappa$(T=273K)   &   1.69   (1.74)$^{(b)}$ &  W/cm/K   \\
                    &  $\kappa$(T=900K)   &   1.21   (1.21)$^{(b)}$           \\
         resistivity  & $\rho_{\rm{FP}}$   &   27.0    (27)$^{(c)}$ & $\mu \Omega$ m/at.fr.  \\
         &  $\rho_{\rm{vac}}$   &   8.11    (7)$^{(d)}$  \\
         constant  & $\frac{c_e v_F^2}{3\Omega_0 T}$ & 1.643$\times 10^{-8}$   & eV/K$^2$/\AA/fs$^2$
    \end{tabular}
    \caption{Parameters fitted to the experimental thermal conductivity as a function of temperature and scattering due to a Frenkel pair in tungsten. 
    We provide fitted parameters suited for a conjugate-gradient relaxed system, for a snapshot from an MD simulation, and for computing the electronic thermal conductivity independently from the phonon contribution.
    Experimental properties given in parentheses: a) ref~\cite{Finnis_PMA1984}, b) ref~\cite{Ho_JPhysChemRefDat1972}, c) ref~\cite{Ullmaier_1991}, d) ref~\cite{Broeders_JNM2004}.
    }
    \label{tab:summary}
\end{table}

\section{Phonon contribution to thermal conductivity}
\label{sec:phonons}

As noted above, the phonon contribution to thermal conductivity is a small fraction of the total for a good conductor, and as it mirrors the trends seen in electronic conductivity the correction due to including a phonon calculation will often be small. 
But it is quite possible to include phonon contributions explicitly, and in this section we will briefly outline how this can be done.

Phonon thermal conductivity can be computed in molecular dynamics using the non-equilibrium M\"uller-Plathe method \cite{MullerPlathe_JCP1997} which matches heat flux to thermal gradients, or using the equilibrium Green-Kubo method \cite{Green_JCP1954,Kubo_JPSJ1957} which uses the velocity autocorrelation function. Both are implemented in LAMMPS. As our low-dose simulated irradiation described below may be sensitive to changes in temperature, we have opted to use the latter method.
 
As noted above, the phonon correction for the electron-impurity scattering coefficient $\sigma_0$ can be neglected. We can refit the values for the electron-phonon and electron-electron scattering coefficients if we assume that a proportion of the experimentally measured thermal conductivity is through phonons. We can compute this phonon contribution using Molecular Dynamics. The phonon conductivity $\kappa_{ph}$ was computed using the Green-Kubo method in the NVT ensemble using a simulation box of 65536 atoms is plotted in figure \ref{fig:diffusivity} - note its small magnitude compared to the total. This value we subtract from the total experimental value, and refit equation \ref{eqn:thermal_conductivity} to the lower electronic-only conductivity. This gives the new fitted parameters $\sigma^{exc}_1 = 1.344\times10^{-4}$ fs$^{-1}$K$^{-1}$ and $\sigma^{exc}_2 = 1.010\times10^{-7}$ fs$^{-1}$K$^{-2}$.

When faced with a new atomic configuration, we can  use the refitted electronic scattering rates $\sigma^{exc}_1$ and $\sigma^{exc}_2$ in equation \ref{eqn:scattering_rate} to compute an electron-only conductivity $\kappa_{el}$, and compute the phonon part $\kappa_{ph}$ afresh using molecular dynamics.
For the defect free system, the sum of the two gives a total conductivity very close to using equation \ref{eqn:scattering_rate} with the original coefficients $\sigma^{MD}_i$ or $\sigma^{CG}_i$.
This demonstrates that for the defect-free system at least, there is little advantage to adding a separate phonon calculation.  

\section{High dose microstructures}
\label{sec:microstructures}
\subsection{MD simulation}

To generate some representative simulated microstructures for this study, we employed a two-step process, described in detail in ref~\cite{Mason_PRM2021}.
First we used the Creation-Relaxation Algorithm (CRA)~\cite{Derlet_PRM2020}, which generates high dose microstructures rapidly, but leaves an excessive number of high energy defects, then we relaxed further with low energy molecular dynamics (MD) cascade simulations~\cite{Gra15,Byg18,Vel17,Gra20}.
The convergence of the results of combined CRA+MD with the results of MD only simulations and their match to other experiments is discussed in ref~\cite{Mason_PRM2021}.

We start with a box of $64\times 64\times 200$ conventional bcc unit cells with a lattice parameter $a_0 = 3.1652 \AA$. The CRA algorithm then selects some atoms at random, and removes them, leaving vacant sites. These are then replaced into random positions, and the simulation cell relaxed using conjugate gradients. We chose LAMMPS and the MNB potential~\cite{Mason_JPCM2017} for the relaxations. During the relaxation, the x- and y- axes were constrained to zero strain, but the z- axis was allowed to relax to zero stress. These elastic boundary conditions are appropriate for simulating an irradiated thin surface layer, constrained by a semi-infinite substrate. This is appropriate for modelling self-ion irradiation in a thick sample~\cite{Mason_PRL2020}. The process of removing and replacing atoms builds up damage, with a canonical measure of the damage given by the ratio of the number of atoms repositioned to the number in the simulation. We displaced 1024 atoms per relaxation, corresponding to $6.25\times 10^{-4}$ canonical displacements per atom (cdpa) per relaxation.

The MD simulations started with the CRA simulations at a range of cdpa values, given in table~\ref{tab:simulation}. These were then strained in the x- and y- directions to the potential's lattice parameter at 300K. The simulation was then thermalized for 20 ps, with a Berendsen thermostat and barostat~\cite{Ber84} to keep zero pressure in the z- direction. The MD simulations were performed using PARCAS~\cite{gha97,nor97,nor95} with the same potential used for the CRA simulations.
Displacement cascades were initiated by shifting the cell randomly in x-, y- and z- directions, maintaining periodic boundary conditions, then giving the central atom 10 keV kinetic energy in a random direction. The cascade was followed with an electronic friction applied to atoms with kinetic energy over 10 eV~\cite{Sand_EPL2013} for 20 ps with a thermostat applied to the border atoms. Finally the simulation was followed for a further 10 ps with a barostat on the z-direction. A new cascade was then initiated.
A canonical dpa level can be associated with these MD simulations by noting the number of vacancies produced per cascade initiated at the initial stages of damage production.  
From the first 40 cascades we estimate a cdpa level $4.1 \times 10^{-6}$ per cascade.

\begin{figure}
    \centering
    \includegraphics[width=1 \linewidth] {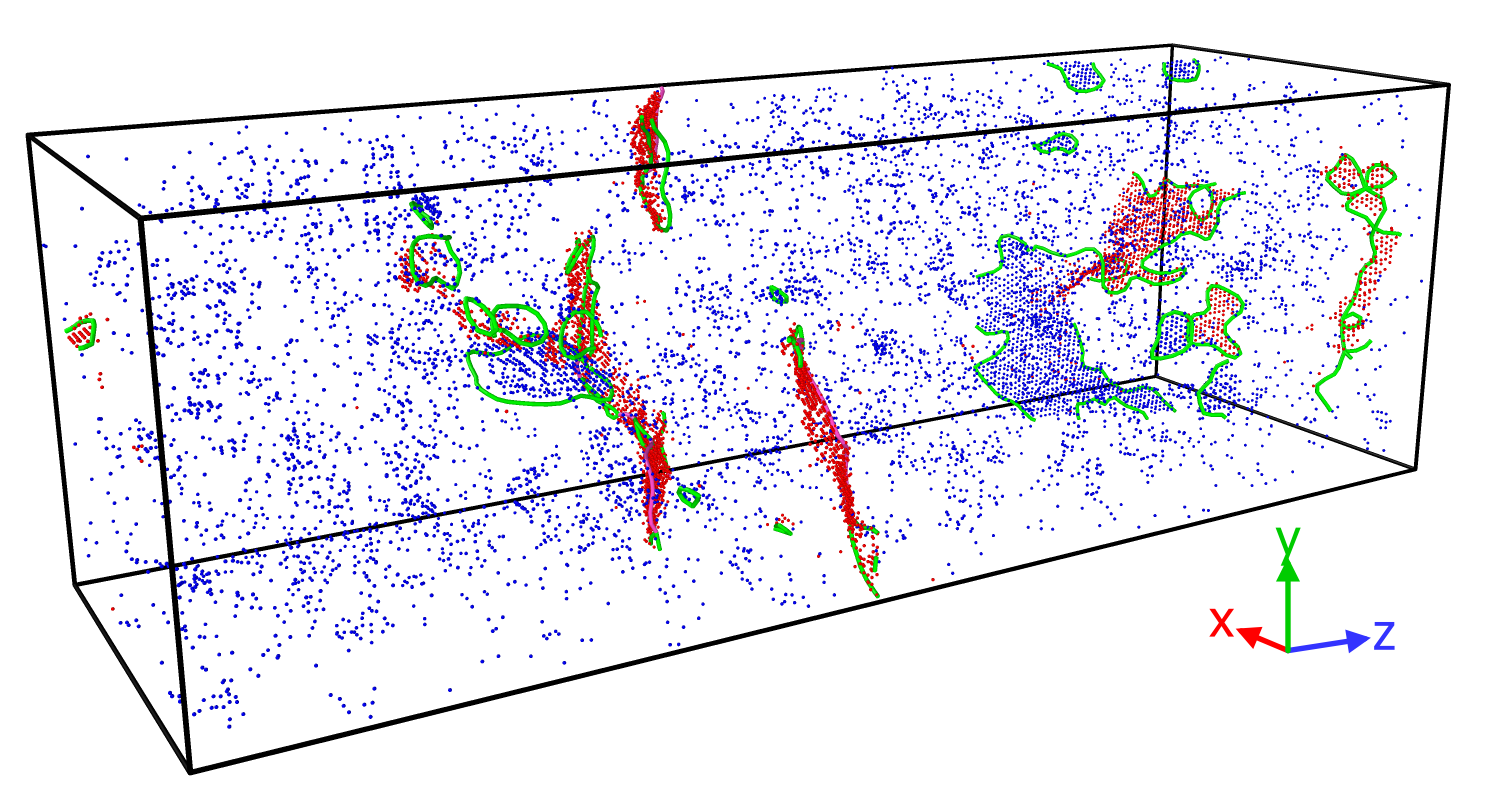}
    \caption{Simulated microstructure at a dose 1.1 cdpa. Dislocation lines with Burgers vectors $1/2\langle 111 \rangle$ (green) and $\langle 100 \rangle$ (pink) generated using DXA\cite{Stukowski_MSMSE2012}. Interstitials (red) and vacancies (blue) generated from Wigner-Seitz cell occupation~\cite{Mason_PRM2021}. Rendered using Ovito~\cite{Stukowski_MSMSE2009}.
    }
    \label{fig:1dpa_microstructure}
\end{figure}

An illustrative simulated microstructure at a dose 1.1 dpa is shown in figure~\ref{fig:1dpa_microstructure}. Note that vacancies are homogeneously dispersed, and dislocation loops of both interstitial and vacancy type can be seen. No isolated crowdions remain.

\begin{table}[htb!]
    \centering
    \begin{tabular}{cccccc}    
        CRA dose    &   MD dose     &   total dose     & $\kappa_{ph}$  & $\kappa_{el}+\kappa_{ph}$ &$\kappa$\\
        (cdpa)      &   (cdpa)      &   (cdpa)        &W/m/K   & W/m/K          & W/m/K         \\        
        \hline
0	    &           0	                    &           0                    &   14.2   & 173.5 &  172.6    \\
0	    &           $4.1\times10^{-5}$	    &           $4.1\times10^{-5}$   &          &       &  169.3    \\
0	    &           $1.63\times10^{-4}$	    &           $1.63\times10^{-4}$  &          &       &  160.3    \\
0	    &           $4.07\times10^{-4}$	    &           $4.07\times10^{-4}$  &          &       &  149.9    \\
0	    &           0.00163	                &           0.00163              &          &       &  132.9    \\
0	    &           0.00407	                &           0.00407              &          &       &  123.6    \\
0	    &           0.00814	                &           0.00814              &          &       &  118.4    \\
0	    &           0.0122	                &           0.0122               &          &       &  116.7    \\
0.00625	&           0.00651	                &           0.0128               &   9.7    & 119.4 &  115.8    \\
0.0188	&           0.00651	                &           0.0253               &   10.0   & 112.3 &  107.6    \\
0.0350	&           0.00651	                &           0.0416               &   8.1    & 103.8 &  100.2    \\
0.0625	&           0.00651	                &           0.0691               &   8.7    & 99.3  &  94.6     \\
0.113	&           0.00651	                &           0.119                &   7.0    & 92.4  &  89.0     \\
0.188	&           0.00651	                &           0.194                &   7.1    & 96.6  &  93.4     \\
0.350	&           0.00651	                &           0.357                &   8.6    & 101.3 &  97.0     \\
0.625	&           0.00651	                &           0.633                &   7.3    & 92.9  &  89.2     \\
1.13	&           0.00651	                &           1.13                 &   8.9    & 100.2 &  95.3     \\
3.00	&           0.00651	                &           3.01                 &   8.0    & 97.9  &  93.8     \\                  
    \end{tabular}                                                                                                 
    \caption{Simulation parameters for generating high dose microstructures, together with the computed thermal conductivity, and the separated out phonon contribution. The error on the computed phonon contribution is order $\pm 0.4$ W/m/K.} 
    \label{tab:simulation}
\end{table}

\subsection{Experimental measurement}
 
Samples of high purity tungsten (99.97 wt\% purity, procured from Plansee) were annealed at 1500C for 24h in vacuum to allow full recrystallization, and then mechanically and electropolished using 0.1\% NaOH solution to produce a mirror finish. 
Ion implantations were then performed at the Helsinki Accelerator Laboratory with 20 MeV W$^{5+}$ ions~\cite{Tikkanen_NIMB2004}.
A summary of the ion fluxes is given in table~\ref{tab:experimental} together with a damage level computed using SRIM (Quick K-P method, assuming threshold displacement energy 68 eV.) These calculations also suggest the peak damage is at a depth 1.25 $\mu$m, falling to near zero at 2 $\mu$m. The peak concentration of injected ions is at $1.7$ $\mu$m.
A full description of the preparation and ion irradiation for these samples is given in ref~\cite{Reza_Acta2020}.
We note that this set of samples has been analysed for other properties, including lattice strain~\cite{Mason_PRL2020} and hardness~\cite{Das_SNApplSci2019}.

\begin{table}[htb!]
    \centering
    \begin{tabular}{cc|c}

       Incident                &   Flux                  &   Damage level     \\
       Fluence                 &                         &   (SRIM)           \\
       (ions/cm$^2$)           &  (ions/cm$^2$/s)        &   dpa              \\
\hline                                                                        
       2.7 $\times 10^{10}$    &  6.24$\times 10^{8}$    &  $1.0\times10^{-4}$           \\
       8.13$\times 10^{10}$    &      "                   &  $3.2\times10^{-4}$           \\
       2.42$\times 10^{11}$    &  3.1-5.0$\times 10^{8}$ &  0.0010            \\
       8.03$\times 10^{11}$    &     "                    &  0.0032            \\
       2.55$\times 10^{12}$    &    "                     &  0.010             \\
       4.61$\times 10^{12}$    &    "                     &  0.018             \\
       8.20$\times 10^{12}$    &    "                     &  0.032             \\
       1.42$\times 10^{13}$    &    "                     &  0.056             \\
       2.54$\times 10^{13}$    &    "                     &  0.10              \\
       8.11$\times 10^{13}$    &   "                      &  0.32              \\
       2.53$\times 10^{14}$    &    "                     &  1.0               \\
       8.10$\times 10^{14}$    &  1.12$\times 10^{11}$   &  3.2               \\
       2.53$\times 10^{15}$    &    "                     &  10.0              \\
       8.10$\times 10^{15}$    &   "                      &  32                \\
             \end{tabular}
    \caption{Fluence and flux of the ion beam used to irradiate the samples. A damage level is computed using SRIM.
    Note that the flux is increased in steps to acheive higher fluences in a reasonable experimental time.}
    \label{tab:experimental}
\end{table}

Thermal diffusivity measurements were made using laser-induced transient grating spectroscopy (TGS)~\cite{Kading_ApplPhysa1995,Hofmann_SciRep2015,Hofmann_MRSBull2019}.
This technique uses crossed, pulsed laser beams ( 0.5 ns duration, $\lambda=532$ nm wavelength, 1 kHz repeat frequency ) to generate a temperature grating at the sample surface. The time-dependent decay of this temperature grating is monitored by diffraction of two continuous wave probe beams that are detected using a fast photodiode connected to an oscilloscope. A detailed description of the experimental setup is provided elsewhere~\cite{Reza_RevSciInstrum2020}.
The thermal diffusivity is then determined from the decay of the diffracted intensity.
A full description of the TGS set up for these measurements can be found in ref~\cite{Reza_Acta2020}.
Calculations suggest the thermal diffusivity measured is dominated by a surface thickness $\sim \lambda_{TGS}/\pi$~\cite{Kading_ApplPhysa1995}, which in this case is 1 $\mu$m and so the measurement reported here is due to the thermal diffusivity changes in the implanted layer.

\section{Results}
\label{sec:results}

\begin{figure}
    \centering
    \includegraphics[width=0.7 \linewidth,angle=-90] {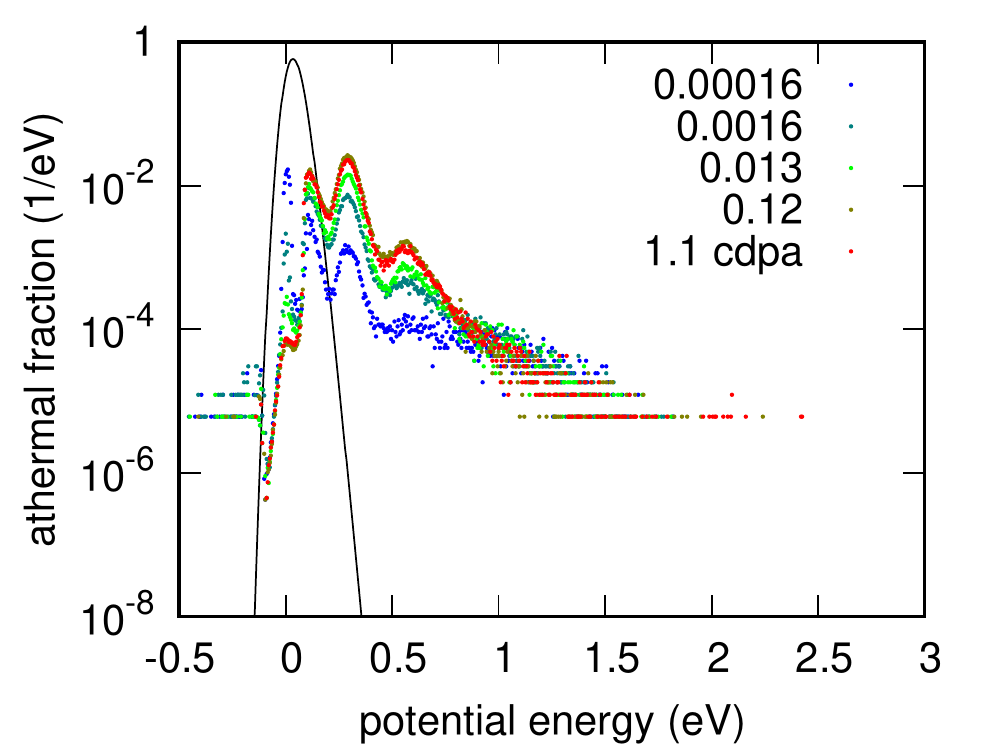}
    \caption{A histogram of potential energies of high dose simulated microstructures.
    The solid line shows the expected fraction of atoms in each bin, normalised so that the area under the curve equals one.
    The symbols show the predicted fraction of non-thermal atoms (equation \ref{eqn:defect_count}, for a range of doses.}
    \label{fig:highDoseSim}
\end{figure}

\begin{figure}
    \centering
    \includegraphics[width=0.7 \linewidth,angle=-90] {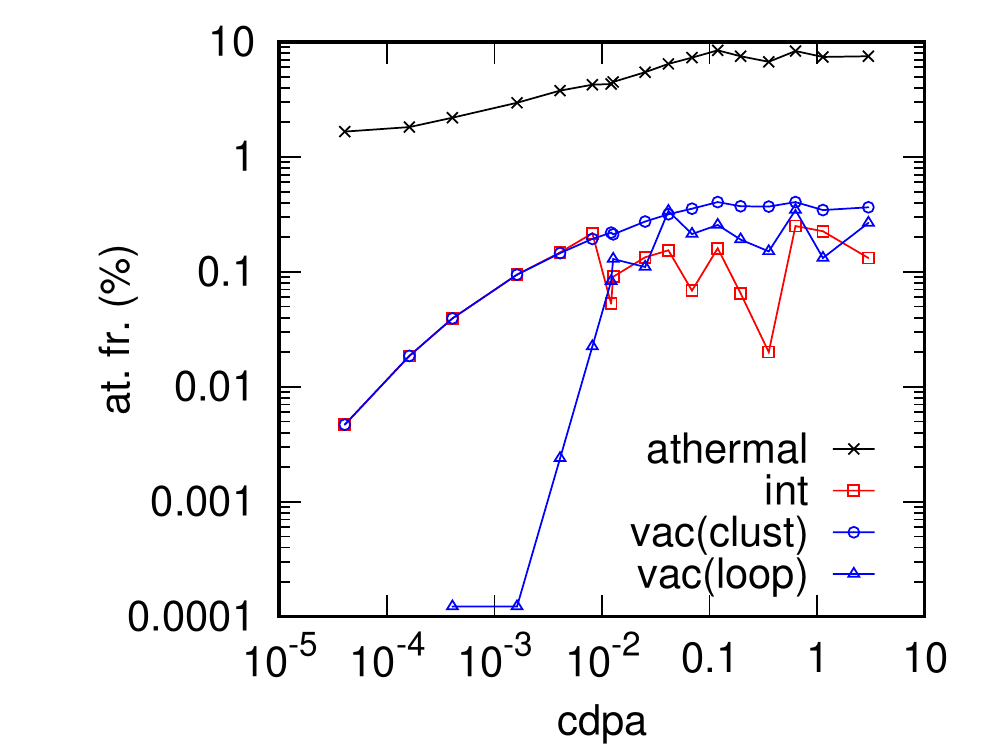}
    \caption{Computed atomic fraction of athermal atoms and defect types for high dose CRA+MD simulations. Interstitials appear mostly as loops, vacancies appear as loops and a homogeneous dispersion of monovacancies and small vacancy clusters.}
    \label{fig:defects_highDoseSim}
\end{figure}

In figure~\ref{fig:highDoseSim} we show the athermal atom count for the relaxed, high-dose microstructure simulations as a histogram binned by potential energy. We can clearly see peaks at $\sim$ 0.3 eV corresponding to vacancies, and over 0.5 eV for interstitials. 
The total athermal atom count for these simulations is plotted in figure~\ref{fig:defects_highDoseSim}.
Note that this is a count of all the atoms which have high energy, and not a count of point defects.
The interstitial and total vacancy count in this figure were computed using a Wigner-Seitz analysis of the occupation of lattice sites, and the vacancy total separated into vacancy clusters and vacancy loops using the method of ref~\cite{Mason_PRM2021}.
We see a saturation of athermal atoms above 0.1 cdpa at about 8\% of the total atom count, while the vacancy concentration saturates at 0.3\%.
This illustrates how a defect in this model is treated as a spatially-diffuse scattering region, and not as the individual point defects.

\begin{figure}
    \centering
    \includegraphics[width=0.7 \linewidth,angle=-90] {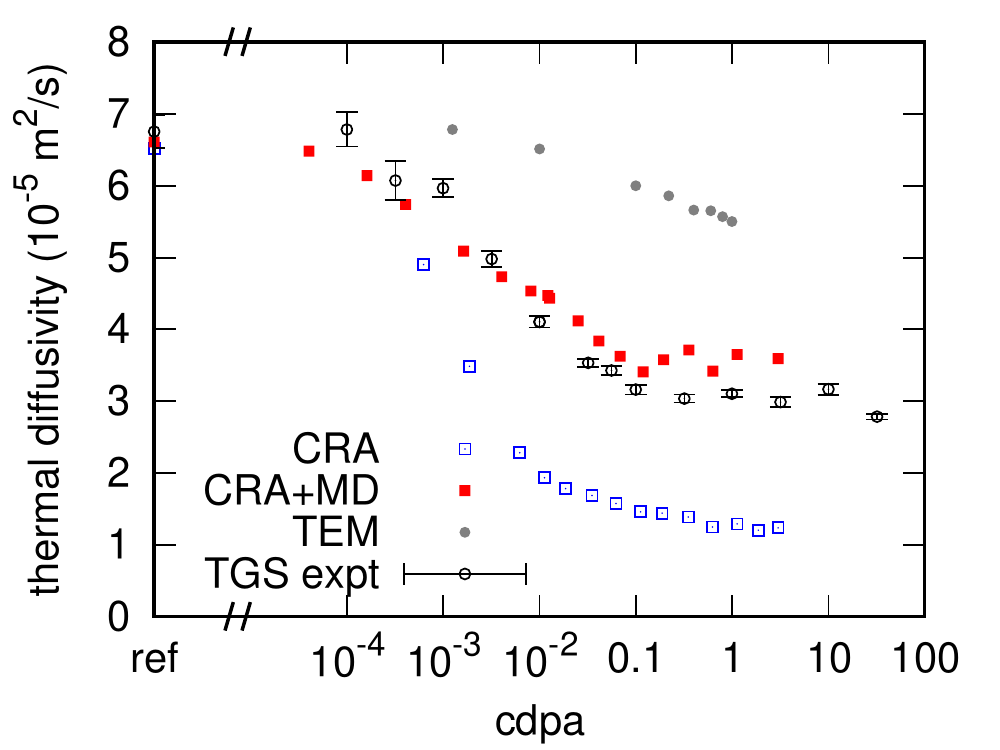}
    \caption{Computed thermal diffusivity of MD simulated microstructures at a range of doses.
    Also shown experimentally measured diffusivity using Transient Grating Spectroscopy (TGS), and an estimate by Reza et al \cite{Reza_Acta2020} of thermal diffusivity due to TEM-visible dislocation loops.
    }
    \label{fig:diffusivity_highDoseSim}
\end{figure}

In figure~\ref{fig:diffusivity_highDoseSim} we show the computed thermal diffusivity for the relaxed high dose microstructure simulations, computed using a single snapshot atomic position file and scattering rates using equation \ref{eqn:total_scattering_rate} parameterized with $\sigma^{MD}_1,\sigma^{MD}_2$.
A separate Green-Kubo calculation for the phonon contribution was performed. We first thermalized the atoms for 100 ps, then sampled the velocity twenty times over 25 ps windows. This was repeated for 25 independent runs, making a total MD sampling time 12.5 ns per data point. We found that the correction was within the size of the data points in figure~\ref{fig:diffusivity_highDoseSim} (order 3-5\%) was made if $\kappa_{el}$ and $\kappa_{ph}$ were computed separately in this way. The values are listed in table \ref{tab:simulation}. We therefore recommend computing thermal conductivity in tungsten using equation \ref{eqn:total_scattering_rate} only, using the values for the scattering rates $\sigma^{MD}_1,\sigma^{MD}_2$ and not using an expensive separate phonon conductivity calculation.

In figure~\ref{fig:diffusivity_highDoseSim} we also include the computed thermal diffusivity for CRA only simulations, with no MD cascade relaxation. 
We see that the unrelaxed CRA-only simulations show the correct general trend seen in the experiment, namely that the thermal diffusivity is significantly reduced as dose increases but saturates over 0.1 dpa. But it is clear that the effect is overestimated. This is an expected consequence of the overestimation of the number of defects generated by the CRA method alone.

Finally, on figure~\ref{fig:diffusivity_highDoseSim} we show an estimate for the thermal diffusivity made by Reza et al.~\cite{Reza_Acta2020} due to TEM visible dislocation loops ($>1.5$ nm diameter). This model uses the area observed in loops in TEM images\cite{Yi_Acta2016} to find a number of interstitial point defects. It is then assumed that each interstitial is paired with a vacancy, and the scattering rate per Frenkel pair is used to turn the observed point defect count into a maximum thermal diffusivity.
As each interstitial is treated as a strong scattering source, even though it may be in the centre of a large dislocation loop and so locally appears as (strained) perfect crystal, this model must overestimate the scattering due to observed defects.
However, this estimate clearly still underestimates the true drop in diffusivity, indicating that visible damage is only a small contributor to the true change in thermal conductivity. In ref~\cite{Reza_Acta2020}, the authors find a better model for the absolute change in thermal diffusivity by assuming defects too small to see follow a power-law distribution~\cite{Sand_EPL2013,Yi_EPL2015}, though can not track the shape of the curve well. 

By contrast to these two estimates, the relaxed CRA+MD cascade simulations show a rate of thermal diffusivity reduction which is a good match to the experiment at doses $<0.1$ dpa, and the saturation level of a 50\% reduction in thermal diffusivity for doses $>0.1$ dpa is also a match. This suggests the level of damage in the relaxed CRA+MD simulations is a good match to experiment at low fluence end where the defect clusters are small, through dislocation network formation at 0.01-0.1 dpa and through to the saturation dose of larger dislocation loop defects seen in figure~\ref{fig:1dpa_microstructure} above 1 dpa.

\section{Conclusion}

In this paper we have used a simple and empirical model for the scattering rate due to a defected atom - stating that the rate should be proportional to the local excess energy difference alone.
This local energy is easy to compute and unambiguous in an MD simulation of a single component system, though we acknowledge that it is not simply accessible to a density functional theory calculation.
However, after this first assumption, we have made no further approximations or experiment-specific parameterizations. 
We developed a simple analytic form for the expected distribution of potential energies, and from this used a statistical method to find the expected number of athermal atoms.
This model can easily be used to post-analyse the output of any single component molecular dynamics simulations.

Though we expect the phonon contribution to thermal conductivity to be small for good conductors, we showed how to incorporate this correction. Both electron and phonon contributions to the conductivity scale with the mean free path of the carriers, a scale set by the defect spacing, so both contributions are reduced as the lattice defects increase. For tungsten, we found the correction due to explicitly separating phonon and electron conductivity to be negligible. We therefore suggest it may be preferable to ignore the phonon contribution entirely for conducting metals, and compute a single scattering rate using equation \ref{eqn:total_scattering_rate} only, with parameters which reproduce observed properties of the thermalized, but undefected crystal.

As electronic thermal transport properties are not accessible to classical empirical potentials, we needed to parameterize the absolute level of the thermal conductivity using established known single crystal experimental data, and we parameterized the scattering rate for the Frenkel pair defect using established electrical resisitivity data.
At high dose the microstructure is one of network dislocations and dislocation loops with a homogeneous background of mono vacancies and small vacancy clusters, and the simulated thermal diffusivity we report is derived from all the athermal atoms.

A natural extension to this model is to include substitutional impurity atoms as point sources of scattering. This was considered in ref~\cite{Hofmann_SciRep2015}, with rhenium atoms in tungsten taken as point sources of impurity scattering. As this approach showed an excellent agreement with experiment, we suggest it should be possible to include impurity atoms in the dilute limit in the present model in a similar way.  

We conclude that our simple model is able to discriminate in a robust manner between undamaged (but strained) crystal, which has only a small contribution to conductivity loss, and highly distorted local environments near dislocation cores and vacancy cages where the scattering should be high. 
As it is fitted to the average scattering rate for a range of atomic environments near Frenkel pairs, correlates with weakly and strongly scattering regions, and correctly deduces the volume fraction of such atomic environments, it is a therefore a good estimator of the average change in thermal diffusivity in highly irradiated simulated microstructures.

\section*{Data Availability}
        
Data and analysis codes are available at https://doi.org/10.5281/zenodo.5724113 .
To obtain further information on the data and models underlying this paper please contact PublicationsManager@ukaea.uk.

\section*{Acknowledgements} 
This work has been carried out within the framework of the EUROfusion Consortium and has received funding from the Euratom research and training programme 2014-2018 and 2019-2020 under grant agreement No 633053 and from the RCUK [grant number EP/T012250/1]. 
AR and FH acknowledge funding from the  European Research Council (ERC) under the European Union’s Horizon 2020
research and innovation programme (grant agreement No. 714697).
The views and opinions expressed herein do not necessarily reflect those of the European Commission.
We would like to thank Max Boleininger for stimulating discussions.
Computer time granted by the IT Center for Science -- CSC -- Finland is gratefully acknowledged.

\bibliography{revised}

\end{document}